# Application of the Kelly Criterion to Ornstein-Uhlenbeck Processes

Yingdong Lv & Bernhard K. Meister

Department of Physics, Renmin University of China
Email:lyd08250@163.com & b_meister@ruc.edu.cn

**Abstract.** In this paper, we study the Kelly criterion in the continuous time framework building on the work of E.O. Thorp and others. The existence of an optimal strategy is proven in a general setting and the corresponding optimal wealth process is found. A simple formula is provided for calculating the optimal portfolio for a set of price processes satisfying some simple conditions. Properties of the optimal investment strategy for assets governed by multiple Ornstein-Uhlenbeck processes are studied. The paper ends with a short discussion of the implications of these ideas for financial markets.

## 1   Introduction

The Kelly Criterion [1], [2] was initially introduced in 1956 to find the optimal betting amount in games with fixed known odds, and was later extended to the field of financial investments by E. O. Thorp and others. The strategy maximizes the entropy and with probability one outperforms any other strategy asymptotically [3]. This approach was recently further developed by Kargin [4], who applied the criterion to a mean-reverting asset process under liquidity and credit constraints.

The Kelly Criterion tells us that the optimal betting fraction is given by *p-q*, if a gambler is faced with a bet, where the probability to double the money is *p* and to lose the initial stake is *q* (*p>q*). The optimal betting fraction maximizes the expected log wealth. The question, why investors should choose to maximize the log wealth, has a simple answer: according to Breiman's theorem [3], it gives the asymptotically optimal pay-out and dominates any other strategy.

In this paper, we start by extending the original idea to the general continuous time framework with *n* correlated assets. Our task is to find the optimal self-financing trading strategy. We will prove that if the market is complete, this optimal self-financing trading strategy always exists. A limited number of applications are discussed in the context of Ornstein-Uhlenbeck processes.

The paper is organized as follows. In section 2 we review the standard assumptions and prove the optimization theorem. The theorem covers both the existence of the optimal trading strategy and the explicit form of the associated optimal investment fraction. In section 3 we apply the theory to a market of *n* correlated assets given by Ornstein-Uhlenbeck mean-reverting processes. The optimal investment strategy is calculated for some representative examples. In the last section we put the results into the financial context and describe some open problems.

## 2   General Theory

In the first section, we will cover the assumptions and the theoretical framework. In 2.2 we will provide the main result, which is contained in Theorem 1.

### 2.1   Basic Assumptions and other Preliminaries

We will use the standard notation and conventions of financial mathematics. In section 2 the basic assumption is that the market is complete and frictionless [5]. Let us further assume that we consider all the processes in the finite time interval 0 to the terminal time T. There exists a probability space $(\Omega, P_T, \mathscr{F}_T)$, on which all of the random variables are constructed, where $\Omega$ is the sample space, $\mathscr{F}_T$ is a $\sigma$-algebra which denotes the information accumulated up to time T and $P_T$ is the spot probability measure [5]. The filtration $\mathscr{F}_t$, $t \in [0,T]$, represents the information accumulated up to time t. The sub-probability space $(\Omega, P_t, \mathscr{F}_t)$ is introduced at time t, where $P_t$ is the restriction of $P_T$ on the filtration $\mathscr{F}_t$.

We assume there are $n+1$ investable assets in the market including the wealth process $B_t$, representing a saving account with value 1 at the initial time 0. We assume $B_t$ follows

$$dB_t = B_t r_t dt \quad , \tag{1}$$

where $r_t$ is the short term rate at time t.

The other $n$ assets in the market are denoted by $S_i(t)$, $t \in [0,T]$, $i = 1, 2, ..., n$, and we define a $n$-dimensional vector by $\mathbf{S}_t = (S_1(t), S_2(t), ..., S_n(t))^T$, where '$T$' represents the transposition of a matrix. Let us define the relative assets price process by $\tilde{\mathbf{S}}_t = \mathbf{S}_t B_t^{-1}$. Let $\phi_0(t)$ denotes the number of units of $B_t$ an investor holds at time t and $\phi_i(t)$, $t \in [0,T]$, $i = 1, 2, ..., n$, denotes the number of units of the $i^{\text{th}}$ asset an investor holds at time $t$. In addition, the $n$-dimensional vector $\phi_t$ is defined as $\phi_t = (\phi_1(t), \phi_2(t), ..., \phi_n(t))^T$. $V_t(\psi)$ is the total value of the portfolio $\psi_t = (\phi_0(t), \phi_t)$. So we have

$$V_t(\psi) = \phi_0(t) B_t + \phi_t \cdot \mathbf{S}_t \quad , \tag{2}$$

where $\phi_t \cdot S_t$ is the inner product of two vectors.

**Definition 1** *A self-financing trading strategy* $\psi_t = (\phi_0(t), \phi_t)$ *is a strategy that satisfies:*

$$dV_t(\psi) = \phi_0(t) dB_t + \phi_t \cdot d\mathbf{S}_t, \quad \forall t \in [0, T] . \quad (3)$$

We assume $V_0(\psi) = 1$.

**Definition 2** *A self-financing trading strategy* $\psi_t = (\phi_0(t), \phi_t)$ *is said to be admissible if and only if*

$$V_t(\psi) \geq 0, \quad P_T \text{ a.s.} \quad \forall t \in [0, T] . \quad (4)$$

$U(x)$, $x \geq 0$, is defined to be a concave function representing the utility of wealth. Here concaveness means

$$U((1-p)x_1 + px_2) \geq (1-p)U(x_1) + pU(x_2), \forall x_2 \geq x_1 \geq 0 \text{ and } 0 \leq p \leq 1 . \quad (5)$$

Further it is assumed that $U(x)$ has a first order derivative for $\forall x \in (0, +\infty)$. The first order derivative at $x = 0$ can be either finite or infinite, and the first order derivative of $U(x)$, $x \geq 0$, is a strictly decreasing function of $x$ with $\lim_{x \to +\infty} U'(x) = 0$. If $U'(0) = +\infty$, let $I(x)$, $x \geq 0$, be the inverse function of $U'(x)$ with $I(0) = +\infty$ and $I(+\infty) = 0$. For $U'(0) = b > 0$, we denote by $I_b(x)$, $x \in [0, b]$, the inverse function of $U'(x)$, with $I(0) = +\infty$. In this case, we define $I(x)$ as

$$I(x) = \begin{cases} I_b(x), & x \in [0, b] \\ 0, & x \in (b, +\infty) \end{cases} . \quad (6)$$

Let us denote by $\mathscr{D}$ the class of all of the admissible self-financing trading strategies. We say a self-financing trading strategy $\psi^* \in \mathscr{D}$ is the optimal trading strategy, if and only if

$$E_{P_T}\left[U(V_T(\psi^*))\right] \geq E_{P_T}\left[U(V_T(\psi))\right], \quad \forall \psi \in \mathscr{D} . \quad (7)$$

Our task is to find an optimal $\psi^* \in \mathscr{D}$, which satisfies eq.7.

### 2.2 The Optimal Strategy

To find the optimal strategy, we will first need to introduce the following lemma..

**Lemma 1** *The function $I(x)$, $x \in [0, +\infty)$ satisfies the following inequality:*

$$U(I(y)) - yI(y) \geq U(c) - yc, \quad \forall y, c \in [0, +\infty) . \quad (8)$$

Proof:

If $I(y) = c$, then eq.8 is obviously satisfied. If $I(y) > c$, then the average growth rate of $U(x)$ from $c$ to $I(y)$ should be larger than the first order derivative of $U(x)$ at $I(y)$, which is given by $U'(I(y))$, since the first order derivative of $U(x)$ is a strictly decreasing function of $x$.

The average growth rate of $U(x)$ from $c$ to $I(y)$ is

$$\frac{U(I(y)) - U(c)}{I(y) - c} .$$

This yields the following inequality

$$\frac{U(I(y)) - U(c)}{I(y) - c} \geq U'(I(y)) = y$$

$$\Rightarrow U(I(y)) - yI(y) \geq U(c) - yc .$$

An almost identical argument can be applied in the case $I(y) < c$. □

Let us define $\tilde{P}_T$ as the martingale of the market and $Z_t = \frac{dP_t}{d\tilde{P}_t}$. Then $(Z_t, \mathscr{F}_t)$ is a $\tilde{P}_T$ martingale [5]. Define $\eta_t^* = y_t Z_t^{-1} B_t^{-1}$, $V_t^* = I(\eta_t^*)$, where we assume $y_t$ is a deterministic function of $t$ and is defined in such a way that $\tilde{V}_t^* = V_t^* B_t^{-1}$ is a $\tilde{P}_T$ martingale. So $y_t$ solves the equation

$$\tilde{E}\left[B_t^{-1} I\left(y_t Z_t^{-1} B_t^{-1}\right)\right] = 1 . \tag{9}$$

The deterministic property of $y_t$ seems contrived, but is necessary for the proof of Proposition 1. As we shall see in section 3, in the case of the log utility function the deterministic function $y_t$ indeed exists and is a constant.

**Proposition 1**   $V_T^* = I(\eta_T^*)$ satisfies the inequality given by eq.7.

Proof:

Let $V_T(\psi)$, $\forall \psi \in \mathscr{D}$, be the wealth process corresponding to a special trading strategy $\psi$, then

$$E_{P_T}\left[U(V_T^*)\right] - E_{P_T}\left[U(V_T(\psi))\right]$$

$$= E_{P_T}\left[\left(U(I(\eta_T^*)) - \eta_T^* I(\eta_T^*)\right) - \left(U(V_T(\psi)) - \eta_T^* V_T(\psi)\right)\right] + E_{P_T}\left[\eta_T^*\left(V_T^* - V_T(\psi)\right)\right] \tag{10}$$

According to lemma 1, the first term of the right hand side of eq.10 is positive. The second term is equal to zero,

$$E_{P_T}\left[\eta_T^*\left(V_T^* - V_T(\psi)\right)\right] = \tilde{E}\left[Z_T\eta_T^*\left(V_T^* - V_T(\psi)\right)\right] = y_T\tilde{E}\left[\tilde{V}_T^* - \tilde{V}_T(\psi)\right] = 0 \quad . \quad (11)$$

The last equality of the above equation is deduced from the fact that both $\tilde{V}_t^*$ and $\tilde{V}_t(\psi)$ are martingales under the martingale measure.

Combining eq.10 and eq.11, we directly get

$$E_{P_T}\left[U\left(V_T^*\right)\right] \geq E_{P_T}\left[U\left(V_T(\psi)\right)\right]$$

□

Proposition 1 only state the fact that $I(\eta_t^*)$ satisfies eq.7. It doesn't necessarily mean that $I(\eta_t^*)$ is the optimal wealth process. We will prove in the following theorem that $I(\eta_t^*)$ is in fact the optimal wealth process.

**Theorem 1** *Given a concave utility function $U(x)$, there exists an optimal self-financing trading strategy $\psi^*$, such that for each time $t \in [0,T]$, the wealth process $V_t(\psi^*)$ of this strategy satisfies*

$$E_{P_T}\left[U\left(V_t(\psi^*)\right)\right] \geq E_{P_T}\left[U\left(V_t(\psi)\right)\right], \quad \forall \psi \in \mathscr{D} \quad .$$

*And the optimal wealth process is given by:* $V_t(\psi^*) = I(\eta_t^*), \quad t \in [0,T]$.

Proof:

Define $\tilde{V}_t^*$ to be $B_t^{-1}I(\eta_t^*)$. For $t=T$, we have $V_T^* = B_T\tilde{V}^* = I(\eta_T^*)$, which represents a general contingent claim in the market. Since the market is complete, the contingent claim $V_T^*$ is attainable. This means there exists a self-financing trading strategy $\psi^*$ such that $\tilde{V}_T^* = V_T(\psi^*)B_t^{-1}$, where $V_t(\psi^*)$ is the wealth process of this self-financing trading strategy. So the relative wealth process $\tilde{V}_t(\psi^*) = V_t(\psi^*)B_t^{-1}$ is a martingale under the martingale measure $\tilde{P}_T$. $\tilde{V}_t^*$ is also a martingale under the martingale measure $\tilde{P}_T$, and we have

$$V_t^* = B_t\tilde{E}\left[\tilde{V}_T^* \mid \mathscr{F}_t\right] = B_t\tilde{E}\left[\tilde{V}_T(\psi^*) \mid \mathscr{F}_t\right] = V_t(\psi^*), \quad \forall t \in [0,T] \quad . \quad (12)$$

Eq.12 shows that $\psi^*$ is a self-financing trading strategy and replicates the optimal wealth process $\tilde{V}_t^* = B_t^{-1}I(\eta_t^*)$. From the combination of $V_t^*$, satisfying eq.7 for any time $t$ before $T$, and eq.12, we can see that $V_t(\psi^*)$ also satisfies eq.7. This proves that the strategy $\psi^*$ is both self-financing and optimal. □

It follows from Proposition 1 and Theorem 1 that the existence of a self-financing trading strategy $\psi^* \in \mathscr{D}$, where the total wealth at a fixed time $T$ is consistent with eq.7, implies that the wealth process $V_t(\psi^*)$ satisfies

$$E_{P_T}\left[U\left(V_t(\psi^*)\right)\right] \geq E_{P_T}\left[U\left(V_t(\psi)\right)\right], \quad \forall \psi \in \mathscr{D} \ .$$

Therefore, an optimal trading strategy for a fixed time $T$ will be optimal for any time before $T$. It follows further that an optimal trading strategy is only based on the information up to time $t$. The optimal trading strategy $\psi_t^*$ is an adapted process with respect to the filtration $\{\mathscr{F}_t, t \in [0, +\infty)\}$. In the next section, we will apply the theorem to the case of a financial market containing a saving account and $n$ correlated assets, whose price processes follow Ornstein-Uhlenbeck mean-reverting processes.

## 3   Implications for Ornstein-Uhlenbeck Processes

In this section we set $r_t = r$ and $S_i(t) = \exp(x_i(t))$, $t \in [0, T]$, $i = 1, 2, ..., n$. Each $x_i(t)$ is governed by

$$dx_i(t) = \left[a_i - b_i x_i(t)\right]dt + \sum_{j=1}^{n}\sigma_{i,j}dW_t^j, \quad i, j = 1, 2, ..., n \ .$$

where $a_i$ is some fixed real number, $b_i > 0$ is some nonnegative real number and $\sigma_{i,j}$ are constants. $\mathbf{W}_t = (W_t^1, W_t^2, ..., W_t^n)^T$ is a standard $n$-dimensional Brownian motion.

Define $\mathbf{a}$ to be the vector $(a_1, a_2, ..., a_n)^T$ ('$T$' is the transposed of a matrix), $\mathbf{b}$ to be the $n \times n$ matrix of the form $\mathbf{b} = \begin{cases} \mathbf{b}_{i,j} = b_i, & i = j \\ \mathbf{b}_{i,j} = 0, & i \neq j \end{cases}$, and $\boldsymbol{\sigma}$ to be the matrix $\boldsymbol{\sigma}_{i,j} = \sigma_{i,j}$, for $1 \leq i, j \leq n$. The matrix $\boldsymbol{\sigma}$ has a non-zero determinant. Then the dynamic equation of $\mathbf{x}_t = (x_1(t), x_2(t), ..., x_n(t))^T$ can be expressed as

$$d\mathbf{x}_t = [\mathbf{a} - \mathbf{b}\mathbf{x}_t]dt + \boldsymbol{\sigma}d\mathbf{W}_t \ . \tag{13}$$

Let $\mathbf{S}_t = (S_1(t), S_2(t), ..., S_n(t))^T$, $\tilde{\mathbf{S}}_t = B_t^{-1}\mathbf{S}_t$.

According to Ito's lemma, the dynamic of $\mathbf{S}_t$ is

$$dS_i(t) = S_i(t)\mu_i(t)dt + S_i(t)\sum_{j=1}^{n}\sigma_{i,j}dW_t^j \ , i = 1, 2, ..., n \ .$$

where $\mu_i(t) = a_i - b_i \log(S_i(t)) + \frac{1}{2}\|\boldsymbol{\sigma}_i\|^2$ and $\boldsymbol{\sigma}_i = (\sigma_{i,1}, \sigma_{i,2}, ..., \sigma_{i,n})^T$.

Define

$$\frac{d\mathrm{P}_T}{d\tilde{\mathrm{P}}_T} = \exp\left(\int_0^T \boldsymbol{\theta}_u \cdot d\tilde{\mathbf{W}}_u - \frac{1}{2}\int_0^T \|\boldsymbol{\theta}_u\|^2\, du\right) \ , \tag{14}$$

where $\boldsymbol{\theta}_u = (\theta_1(u), \theta_2(u), ..., \theta_n(u))^T$ is a *n*-dimensional adapted stochastic process and $\|\boldsymbol{\theta}_u\| = \sqrt{\theta_1^2(u) + ... + \theta_n^2(u)}$ is the Euclidean vector norm, and $\tilde{\mathbf{W}}_t$ a Girsanov transformed Brownian motion, i.e. $\mathbf{W}_t = \tilde{\mathbf{W}}_t - \int_0^t \boldsymbol{\theta}_u du$.

Then under $\tilde{\mathrm{P}}_T$, $\tilde{\mathbf{S}}_t$ it follows

$$d\tilde{S}_i(t) = \tilde{S}_i(t)\left[\mu_i(t) - r - \sum_{j=1}^{n}\sigma_{i,j}\theta_j(t)\right]dt + \tilde{S}_i(t)\sum_{j=1}^{n}\sigma_{i,j}d\tilde{W}_t^j \ .$$

Define $c_i(t) = \mu_i(t) - r$ and in vector form $\mathbf{c}_t = (c_1(t), c_2(t), ..., c_n(t))^T$. If $\boldsymbol{\theta}_t$ solves

$$\boldsymbol{\sigma}\boldsymbol{\theta}_t = \mathbf{c}_t \Rightarrow \boldsymbol{\theta}_t = \boldsymbol{\sigma}^{-1}\mathbf{c}_t \ , \tag{15}$$

then under $\tilde{\mathrm{P}}_T$ it follows that

$$d\tilde{\mathbf{S}}_t = \tilde{\mathscr{S}}_t \boldsymbol{\sigma} d\tilde{\mathbf{W}}_t \ , \tag{16}$$

where the matrix $\tilde{\mathscr{S}}_t$ is defined as: $\tilde{\mathscr{S}}_t = \begin{cases} \tilde{\mathscr{S}}_{i,j}(t) = \tilde{S}_i(t), & i = j \\ \tilde{\mathscr{S}}_{i,j}(t) = 0 & , i \neq j \end{cases}$. $\tilde{\mathbf{S}}_t$ is a martingale under $\tilde{\mathrm{P}}_T$.

To apply theorem 1, we need first to prove the completeness of the market price processes under consideration. The next lemma tells us that indeed the market is complete. The proof is given in an earlier presentation [7].

**Lemma 2**   *The mean-reverting market given above is complete.*

In case $U(x) = \log(x)$, we find $I(x) = 1/x$. Using eq.9, we can show $y_t = 1$. Then the optimal discounted wealth process is

$$\tilde{V}_t^* = Z_t = \exp\left(\int_0^t \boldsymbol{\theta}_u \cdot d\tilde{\mathbf{W}}_u - \frac{1}{2}\int_0^t \|\boldsymbol{\theta}_u\|^2\, du\right).$$

Now we are in a position to derive a general result for Ornstein-Uhlenbeck processes.

*Application of the Kelly Criterion to Ornstein-Uhlenbeck Processes*

**Theorem 2** *The optimal trading strategy $\psi_t^* = \left(\phi_0^*(t), \phi_t^*\right)$ is given by:*

$$\phi_0^*(t) = B_t^{-1}V_t^*\left(1 - B_t^{-1}\boldsymbol{\theta}_t^T\boldsymbol{\lambda}_t\mathbf{S}_t\right), \ \phi_i^*(t) = B_t^{-1}V_t^*\sum_{j=1}^n \theta_j(t)\lambda_{j,i}(t), \ i = 1,2,...,n \quad . \quad (17)$$

where $\boldsymbol{\lambda}_t = \left(\tilde{\mathscr{S}}_t\boldsymbol{\sigma}\right)^{-1}$.

Proof:
First, we can show immediately

$$V_t^* = \phi_t^* \cdot \mathbf{S}_t + \phi_0^*(t)B_t \quad ,$$

where $V_t^*$ is the optimal wealth process given by

$$V_t^* = B_t \exp\left(\int_0^t \boldsymbol{\theta}_u \cdot d\tilde{\mathbf{W}}_u - \frac{1}{2}\int_0^t \|\boldsymbol{\theta}_u\|^2 \, du\right) \quad .$$

Using Ito's lemma for $\tilde{V}_t^*$, we get

$$d\tilde{V}_t^* = dZ_t = \tilde{V}_t^* \boldsymbol{\theta}_t^T d\tilde{\mathbf{W}}_t \quad .$$

From eq.16 we know that

$$d\tilde{\mathbf{W}}_t = \boldsymbol{\lambda}_t d\tilde{\mathbf{S}}_t \quad .$$

Combining the above two equations, we have

$$d\tilde{V}_t^* = \tilde{V}_t^*\boldsymbol{\theta}_t^T\boldsymbol{\lambda}_t d\tilde{\mathbf{S}}_t = B_t^{-1}\tilde{V}_t^*\left(\boldsymbol{\theta}_t^T\boldsymbol{\lambda}_t d\mathbf{S}_t - r\boldsymbol{\theta}_t^T\boldsymbol{\lambda}_t\mathbf{S}_t dt\right) \quad , \tag{18}$$

and

$$d\tilde{V}_t^* = B_t^{-1}dV_t^* - rB_t^{-1}V_t^* dt \quad . \tag{19}$$

Combining eq.18 and eq.19, we arrive at

$$dV_t^* = V_t^*\left[B_t^{-1}\boldsymbol{\theta}_t^T\boldsymbol{\lambda}_t d\mathbf{S}_t + B_t^{-1}\left(1 - B_t^{-1}\boldsymbol{\theta}_t^T\boldsymbol{\lambda}_t\mathbf{S}_t\right)dB_t\right] = \phi_t^* \cdot d\mathbf{S}_t + \phi_0^*(t)dB_t \quad . \tag{20}$$

Eq.20 shows directly that $\psi_t^* = \left(\phi_0^*(t), \phi_t^*\right)$ given by eq.17 is the optimal self-financing trading strategy. $\square$

The optimal fraction vector $\mathbf{f}_t^* = \left(f_1^*(t), f_2^*(t),...,f_n^*(t)\right)^T$ is composed of the individual $f_i^*(t)$, e.g. $\phi_i^*(t)S_i^*(t)/V_t^*$. By simple calculations based on Theorem 2, we have

$$\mathbf{f}_t^* = \mathbf{R}^{-1}\mathbf{c}_t, \tag{21}$$

where $\mathbf{R} = \boldsymbol{\sigma}\boldsymbol{\sigma}^T$ is a symmetric matrix and called correlation matrix. We will show in a separate paper that the matrix $\mathbf{R}$ denotes the correlations of the yield rates, e.g. the correlation of the $i^{th}$ and $j^{th}$ assets is a deterministic function of $\boldsymbol{\sigma}_i \cdot \boldsymbol{\sigma}_j$. If the standard inverse of the volatility matrix does not exist, then one can resort to the generalized Moore-Penrose inverse to obtain a related result for the optimal investment fractions in markets without arbitrage.

Another derivation of the optimal fraction can be based on the function

$$F(\mathbf{x}) = \mathbf{c}_t^T \mathbf{x} - \frac{1}{2} \mathbf{x}^T \mathbf{R} \mathbf{x}, \qquad \forall \mathbf{x} \in \mathbf{R}^n \quad , \tag{22}$$

linked to the mean-variance approach, since the optimal fraction given by eq.21 is the maximum of the function $F$. This indicates the close relationship between the utility maximization and the mean-variance method.

In the special case where the market is composed of only one stock, by eq.21, the optimal fraction is $f_t^* = (\mu_t - r)/\sigma^2$, where $\mu_t = a - b\log(S(t)) + 0.5\sigma^2$. Fig. 1 shows a sample path for the stock process and the associated optimal investment fraction and wealth process. As an aside, if one assumes zero interest rates, than the sensitivity of the optimal fraction to a percentage estimation error in the drift $\mu$ is twice the negative of a similar error in $\sigma$, e.g. a *1% overestimation in volatility has approximately the same impact as an underestimation of the drift by 2%*.

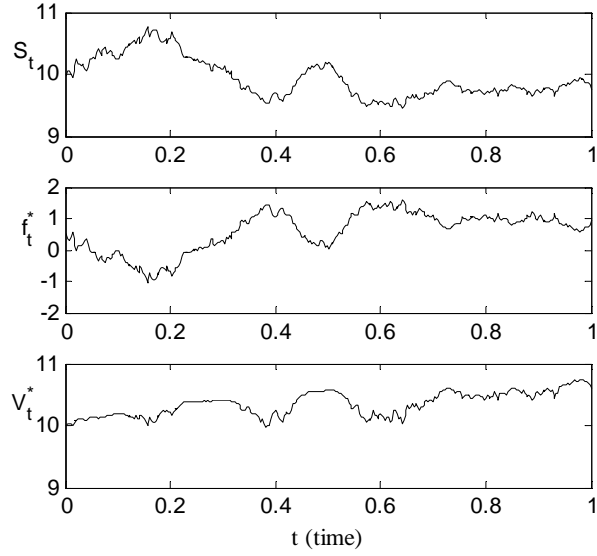

**Fig. 1.** Simulation of the stock price process, the corresponding optimal strategy $f_t^*$, and the wealth process $V_t^*$ with parameters a=0.5, b=0.2, $\sigma$ =0.1, r=0.03, $S_0$ =10 and $V_0$ =10.

Besides the sensitivity to estimation errors, it would be interesting to analyze the impact of the correlation matrix on the optimal trading strategy. Positive correlation has a tendency to reduce the number of 'independent' assets and forces investors to reduce leverage, e.g. the sum of the absolute values of the investment fractions is smaller.

Next, we study a special case where the assets have local correlations. The different assets only correlate to the neighboring assets but have no correlation to the rest assets. Let's set the risk-free rate to zero and the volatility matrix to be

$$\boldsymbol{\sigma} = \begin{bmatrix} \sigma & \sigma & 0 & \cdots & 0 \\ 0 & \sigma & \sigma & \ddots & \vdots \\ 0 & 0 & \ddots & \ddots & 0 \\ \vdots & \ddots & \ddots & \ddots & \sigma \\ 0 & 0 & 0 & \cdots & \sigma \end{bmatrix}_{n \times n} , n \geq 2 \ .$$

In this case, we will only study the large time limit. Let us denote by $\bar{f}_S^*(\infty)$

$$\bar{f}_S^*(\infty) := \lim_{t \to +\infty} E_{P_T}\left[\sum_{i=1}^n f_i^*(t)\right] \ . \tag{23}$$

After some simple manipulations we get

$$\bar{f}_S^*(\infty) = \begin{cases} \dfrac{n+1}{4} & \text{for n odd} \\ \dfrac{n}{4} & \text{for n even} \end{cases} . \tag{24}$$

For a fixed odd integer *n*, the limit of the expected total fraction is (*n*+1)/4, which is identical to the value for the next even number.

Now, let us investigate another correlation structure where the assets have global correlations. As a simple example the volatility matrix is chosen as

$$\boldsymbol{\sigma} = \sigma \begin{bmatrix} 1 & 0 & 0 & \cdots & 0 \\ 1 & 1 & 0 & \cdots & 0 \\ 1 & 1 & \ddots & \ddots & \vdots \\ \vdots & \vdots & \ddots & 1 & 0 \\ 1 & 1 & \cdots & 1 & 1 \end{bmatrix}_{n \times n} , \ n \geq 2 \ ,$$

and the corresponding inverse matrix is

$$\boldsymbol{\sigma}^{-1} = \sigma^{-1} \begin{bmatrix} 1 & 0 & 0 & \cdots & 0 \\ -1 & 1 & 0 & \cdots & 0 \\ 0 & -1 & \ddots & \ddots & \vdots \\ \vdots & \vdots & \ddots & 1 & 0 \\ 0 & 0 & \cdots & -1 & 1 \end{bmatrix}_{n \times n} \ .$$

Thus the optimal fraction $f_i^*(t)$ is given by

$$f_i^*(t) = \begin{cases} c_1(t) - (c_2(t) - c_1(t)) & i=1 \\ (c_i(t) - c_{i-1}(t)) - (c_{i+1}(t) - c_i(t)) & i=2,3,\ldots,(n-1) \\ c_n(t) - c_{n-1}(t) & i=n \end{cases} . \quad (25)$$

The total optimal fraction is

$$f^*(t) = \sum_{i=1}^{n} f_i^*(t) = c_1(t) = \frac{\mu_1(t) - r}{\sigma^2} . \quad (26)$$

The total fraction here is equal to the optimal fraction in another market containing only the first asset. This surprising result is partly due to the fact that in the multi-dimensional case the investment fractions are likely to have both positive and negative signs. The expected total optimal fraction is

$$\overline{f}^*(t) = E[f^*(t)] = \overline{c}_1(t) = \frac{[a_1 - b_1 \log(S_1(0))]e^{-b_1 t} + \frac{1}{2}\sigma^2 - r}{\sigma^2} . \quad (27)$$

As time approaches infinity, the following limit is reached:

$$\overline{f}^*(\infty) = \lim_{t \to +\infty} \overline{f}^*(t) = \begin{cases} \dfrac{a_1 + \frac{1}{2}\sigma^2 - r}{\sigma^2} & \text{for } b_1 = 0 \\ \dfrac{\frac{1}{2}\sigma^2 - r}{\sigma^2} & \text{for } b_1 \neq 0 \end{cases} . \quad (28)$$

For additional examples, in particular in higher dimensions, we refer the reader to [7].

## 4 Conclusions

In the earlier sections we presented a discussion of the Kelly criterion in the continuous-time framework. The main theorem shows that in a complete market there exists an optimal self-financing trading strategy that maximizes the logarithmic utility function. The optimal investment fractions were explicitly calculated.

One general implication of the Kelly's criterion is maybe worth mentioning. It follows from Breiman's theorem [2], which shows that a logarithmic utility maximizer outperforms with probability one in the long run any substantially different trading strategy. This theorem has surprising consequences, for example it has spanned a smallish field called 'evolutionary finance' [8]. According to evolutionary finance 'natural selection' should favor agents with log utility. Such agents maximize the growth rate of their wealth with probability one, and thus dominate eventually the market. The stark claim is that either the investor maximizes utility or is marginalized. The authors are doubtful, if such a strong claim is justified, since only in the long time limit does the utility maximizer almost surely outperform. In the real world, where one has multiple independent agents and frequent paradigm shifts, maybe an even more aggressive strategy is warranted. Being 'overinvested' can be 'superior' (lower utility, but higher winning probability) in the short term. Even in the medium term the log maximizer has difficulties to outperform, if many independent agents exist. This could be a partial explanation for the regular crisis in financial markets, e.g. investors, who seek a short-term competitive advantage, invest over-aggressively. This is on top of the significant inherent volatility of utility maximization. A proper understanding of the impact of the Kelly criterion on the optimal behavior of individual agents is the precursor to consistent multi-agent modeling.

As a speculative aside, maybe utility maximization has a role in the study of the punctuated equilibrium observed in the evolutionary history of the earth, since utility maximization could provide a potential explanation without necessarily having to resort to external causes like asteroid impacts or volcanic eruptions for rare widespread extinction events.

A further interesting aside is to relate Kelly's criteria to market bubbles. A bubble is sometimes defined as a self-reinforcing dynamic associated with an excessive increase in asset prices followed by a sudden collapse. One simple and generic way to achieve a bubble is to find a reinforcing process that reduces the volatility $\sigma$, which would increase the optimal leverage as perceived by the investors, since the optimal Kelly fraction is given by $(\mu_t - r)/\sigma^2$. Next a description of a possible self-reinforcing process: In a number of investment strategies investors are short volatility. This selling of implied volatility can lead to a decrease of implied volatility due to supply and demand imbalances, which leads on the one hand to an increase in the optimal leverage and on the other hand to mark-to-market profits for the short volatility positions. As volatility decreases the optimal position size, e.g. leverage, increases, putting additional pressure on implied volatilities. This process continues until the friction associated with obtaining additional leverage stops the process. This 'virtuous circle' is then replaced by a 'vicious circle', since the extremal point is unstable, as implied volatility increases and leverage decreases. Here we do not discuss realized and implied volatility separately, since they are positively correlated and for the qualitative description presented here a dampening process for either type of volatility is sufficient. The analysis sketched out above can be extended to show that many bubbles, i.e. credit and stock market bubble, are driven by changes in the optimal leverage ratio as derived from the Kelly criterion.

Due the space limitations, we are not able to provide even a brief description of the application of the result in the area of statistical arbitrage. The present discussions are based on the continuous time framework, but realistic markets have an inherent discreteness. Furthermore there are different types of frictions, e.g. transaction cost, bid-offer spreads and liquidity constraints, which impose portfolio readjustment frequency restrictions. Not all of those influences are small and can be neglected. In an earlier presentation [7] correction terms for reducing the investment fractions were explicitly calculated. It would be of interest to give a comprehensive analysis of the impact of the different types of frictions for statistical arbitrage strategies. This will be done by the authors in a separate paper.

In conclusion this article gives a quantitative insight into the trade-off between risk and return as diversification opportunities are added, correlation structure changed, and other constraints modified.

**Acknowledgements.** YL expresses his thanks to one of his dearest friends, Huiqing Chen, for her kind encouragement and support. BKM acknowledges support from the NSF of China and multiple informative discussions with DC Brody and O Peters.